\documentclass{article}
\usepackage[english]{babel}
\usepackage{amsthm}
\usepackage{amsmath}
\usepackage{amsfonts}
\usepackage{babel}
\usepackage{makeidx}
\usepackage{amssymb}
\usepackage{color}

\title{A comment on paper of Kim et al. on mechanisms of hysteresis in human brain networks: comparing 
with theoretical  $m$-adic model }
\author{Giuseppe Iurato and Andrei Khrennikov\\International Center for Mathematical Modeling\\in Physics, Engineering,
Economics and Cognitive Science\\Linnaeus University, S-35195, V\"axj\"o, Sweden}

\date{}
\begin{document}

\maketitle

\abstract{This comment is aimed to point out that the recent work due to Kim, et al. in which the clinical and experiential assessment of a brain network model suggests that
asymmetry of synchronization suppression is the key mechanism of hysteresis has coupling 
with our theoretical  hysteresis model of unconscious-conscious interconnection based on  dynamics
on $m$-adic trees.}

\section{Introduction}
This is a comment on  the recent work \cite{kmml}. In this paper, the clinical and experiential assessment of a brain network model suggests that asymmetry of synchronization suppression is the key mechanism of hysteresis
observed during loss and recovery of consciousness in general anesthesia.  This study  has indirectly provided empirical
confirmation of the theoretical model outlined in \cite{iukh}  based on a possible implementation of an
hysteretic pattern into a formal model of unconscious-conscious interconnection worked out on the basis of
representations of mental entities by $m$-adic numbers.

One of the main assumptions done by the authors of \cite{kmml}, is that (physical) hysteresis (of their brain network model took into
account) observed during anesthetic state transitions shares the same underlying mechanism as that observed in
non-biological networks. This makes licit to put into comparative relations \cite{iukh} and \cite{kmml}.

\section{$M$-adic (ultrametric) model of conscious-unconscious interrelation}

First, rigorous attempts to formalize, through
$p$-adic mathematics, the construct pair conscious-unconscious of psychology have been undertaken by Andrei Yu.
Khrennikov since the late 1990s (\cite{D}, \cite{khr0}-\cite{khr8}). This formalization via $p$-adic analysis
was based on the use of concepts, tools and techniques drawn from dynamical systems theory and this route is
very promising. One of the central points of this theoretical framework, which lays out the basic concepts and
notions of psychology and psychoanalysis, is the use of $p$-adic dynamical systems
and related theory, thanks to which it has been possible to take into account the chief elements of Freudian
psychoanalysis, among which the crucial relationships conscious-unconscious, which may be formalized through
discrete dynamical system theory and represent the nodal points of the whole psychoanalytic framework.

Mathematically, it is fruitful to proceed with the fields of $p$-adic numbers, where $p>1$ is a prime number. These fields play an
important role in theoretical physics, string theory, quantum mechanics and field theory, cosmology -- see,
e.g., \cite{pp2}, \cite{drag}, \cite{dra2} for recent reviews. However, in cognitive and psychological
applications there are no reasons to restrict models to prime number bases. It is more natural to work with the
rings of $p$-adic numbers, where $p>1$ is an arbitrary natural number. In general, the language of ultrametric
spaces covers completely tree-like representations of information in cognitive studies and psychology
(\cite{khr4}). However, up to now not so much has been done on general ultrametric spaces. Finally, we also
remark that methods of $p$-adic and more generally ultrametric analysis, have been used in modeling cognition
and unconscious processing of information by R. Lauro-Grotto (\cite{S1}) and F. Murtagh (\cite{S2}-\cite{S5}).

Therefore, the psychological construct pair conscious-unconscious, say $\mathcal{C}-\mathcal{UC}$, is the
keystone of every formalization attempt of psychoanalysis. In \cite{iukh}, the authors have simply taken into
account a first elementary formal model of hysteretic phenomena (regarding physical context), implemented into
the $p$-adic dynamical model of the $\mathcal{C}-\mathcal{UC}$ pair. In doing so, the authors of \cite{iukh}
have tried to use hysteretic phenomena (belonging to physics) to analogically transfer memory retaining effects
into the phenomenology involved in the pair $\mathcal{C}-\mathcal{UC}$. Indeed, hysteretic effects have been
considered in attempts to mechanically formalize memory features of implicit memories of neurophysiology
(\cite{frack}, \cite{lat}), so the authors of \cite{iukh} have thought to extend this idea to
$\mathcal{C}-\mathcal{UC}$ pair, trying to shed light upon a formal issue raised by the $m$-adic dynamical
model. The model outlined in \cite{iukh} has been then applied to formalize other aspects of human psyche
(\cite{iukhmu}) as well as to deduce a $p$-adic version of the Weber-Fechner law (\cite{iur1}) and some of its
possible applications to economics and sociology (\cite{iur2}).

\section{Hysteresis}

Hysteresis has a large range phenomenology, and may be understood from either the psychological and the physical
standpoint. A possible conception of hysteresis belonging to psychological context may be drawn from the {\it
APA Dictionary of Psychology} which defines hysteresis as an effect in which the perception of a stimulus is
influenced by any other stimulus immediately preceding such a perception. It can be detected, for instance, in
experiments making successive changes to a certain stimulus which is varying along some dimension, hence asking
to the participant to describe her or his perception. When such values along the given dimension are steadily
increased, then it will be reached a point in which the participant will begin to place the related percept into
a different category (e.g., a sound is loud rather than quiet\footnote{This just resembles that typical
phenomenology involved in sound experiences called into question in explaining Weber-Fechner law
(\cite{iur1}).}), but, when values along the dimension are decreased, then the crossover point will occur at a
different point along such a dimension. In particular, in vision, hysteresis may stand out with the tendency for
a perceptual state to persist under gradually changing conditions: this is, for example, the case when
stereoscopic fusion may persist, so producing the appearance of depth even when binocular disparity (i.e., the
slight difference between the right and left retinal images) between the two images becomes so great that they
would normally not be able to be merged together.

This last phenomenology of hysteresis (to be meant according to psychology) related to vision may be also
correlated analogically (\cite{akot}) with certain aspects of the physical phenomenology discussed first in
\cite{khr9}, and dealing with conscious-unconscious visual recognition, hence reconsidered in \cite{akot} where
the authors have then pointed out the possible analogical identification of hysteresis effects in visual
recognition experiments performed in \cite{akoty}. Indeed, in such a context, H. von Helmholtz unconscious
inferences, which play a crucial role in the passage from sensation to perception, are considered in relation to
a quantum-like pattern of sensation-perception dynamics -- quantically treated, in that not based on classical
logics -- so providing a concrete model for unconscious and consciousness processing of information and their
interaction. To be precise, in the cognitive modeling worked out in \cite{khr9} and \cite{akot}, if $S$
represents the unconscious information processing and $S'$ the conscious one, then, in the concrete instance of
von Helmholtz's unconscious inference, $S$ represents just the processing of sensation (its unconscious nature
having been emphasized as early by Hermann von Helmholtz) and $S'$ represents processing of perception-conscious
representation of sensation. The related experiment performed in \cite{akoty}, then theoretically analyzed in
\cite{khr9} and \cite{akot}, concerned the bistable perception (of the type $S\rightarrow S'$) of the rotation
of an ambiguous figure (i.e., the {\it Schr\"{o}der stair}), which turned out to be different, for each of the
three groups of persons chosen to form statistical test samples, due to the diversity of data's contextuality
(suitably treatable just by quantum formalism) entailing optical illusions affected by memory biases, and put
into relation with hysteresis effects in \cite{akot}.

On the other hand, following \cite{kmml}, there already existed a wide literature on computational biology works
which, since the late of 1990s and the beginnings of 2000s, have put attention to possible hysteresis phenomena
(to be meant according to physics and network systems) occurring in a large-scale brain network modelled with
simple oscillatory patterns, in particular during state transitions of consciousness and unconsciousness (like
in general anesthesia and sleep), with hysteresis observed during the loss and recovery of consciousness
mediated by patterns of synchronization meant, according to general network systems, as a pathway discontinuous
transition between incoherent (unconsciousness) and synchronized (consciousness) states of a network\footnote{As
authors themselves point out in \cite{kmml}, consciousness and unconsciousness cannot be trivially reduced to,
respectively, synchronized and incoherent networks, as it is temporal coordination, rather than synchrony, to be
critical for consciousness. This is also in agreement with (and provides partial empirical evidence to) what we
have stated in \cite{iukhmu}.} that is, the asymmetry between the synchronization and desynchronization paths is
just the key network mechanism of hysteresis. The decreasing/increasing of long-range network synchronization is
considered as a basic neural mechanism during the loss/recovery of consciousness\footnote{Indeed, in
\cite{kmml}, it has been ascertained that functional brain networks become more modular during general
anesthesia, in that coordinated and synchronized interactions across the cerebral cortex break down. So, for the
aims of large-scale modeling, authors of \cite{kmml} assume that the conscious brain will have, in aggregate,
more synchronized interactions with respect to the unconscious brain.}. Furthermore, network mechanism of
hysteresis is not as a regional brain activity but rather is a globally conceived mechanism (\cite{kmml}). This
is an remarkable outcome as it proves that (physical) hysteresis is a phenomenon concerning the general psychic
mechanisms of human brain. In particular, in \cite{kmml}, it has been proved that hysteresis occurs above all
during state transitions around a lower lever of consciousness. This justifies the theoretical implementation of
a formal model of hysteretic phenomena (regarding physical context) into the $p$-adic dynamical model of the
$\mathcal{C}-\mathcal{UC}$ pair, as done in \cite{iukh}, where the authors have supposed that hysteresis
mechanism roles functionally unconscious realm\footnote{This is in agreement with the hypothesis, assumed in
\cite{kmml}, for which a stronger anesthetics (hence, a deeper unconscious state) induces a larger hysteresis.}
and the related consciousness processes coming from it.

\end{document}